\documentclass[prb,twocolumn,showpacs,amsmath,amssymb]{revtex4}
\usepackage{graphicx}   
\usepackage{dcolumn}    
\usepackage{bm}         
\newcommand{\rd}{{\rm d}}
\newcommand{\re}{{\rm e}}
\newcommand{\ri}{{\rm i}}

\newcommand{\bra}[1]{{\langle \left.#1\right|}}
\newcommand{\ket}[1]{{\left|#1\right.\rangle}}  

\begin{document}


\title[Control of transfer]{Coherent control of population transfer 
between communicating defects}

\author{Christoph Weiss}
   \email{weiss@theorie.physik.uni-oldenburg.de}
   \affiliation{Institut f\"ur Physik, Carl von Ossietzky Universit\"at,
                D-26111 Oldenburg, Germany}

\date{3 January 2006}

\begin{abstract}
Population transfer between two identical, communicating defects in a
one-dimensional tight-binding lattice can be systematically controlled
by external time-periodic forcing. Employing a force with slowly changing 
amplitude, the time it takes to transfer a particle from one defect to 
the other can be altered over several orders of magnitude. An analytical 
expression is derived which shows how the forcing effectively changes the 
energy splitting between the defect states, and numerical model calculations 
illustrate the possibility of coherent control of the transfer.  
\end{abstract}


\maketitle

\section{Introduction}

The possibility to control the state of a quantum system by external
forcing, or even to create desired target states by judiciously designed
electromagnetic pulses, is of substantial conceptual and practical interest. 
In this paper, a specific scenario is investigated which emerges in 
one-dimensional tight-binding lattices under the influence of periodic 
forcing, such as laser-irradiated polymers, or semiconductor superlattices
interacting with far-infrared radiation. If the on-site energy is 
deliberately modified in an identical manner at two sites of the otherwise 
regular lattice, two defect states emerge which ``communicate'' if they
are not too far apart, {\em i.e.\/}, they possess localized eigenstates
connecting both defects. It will be shown that the degree of communication,
expressed by the splitting of the energies of the two defect states, can
be varied over several orders of magnitude if the system is subjected
to strong periodic forcing. From this follows the possibility to achieve
well-controlled population transfer from one defect to the other, if
the forcing amplitude is suitably shaped.

The paper relies on the framework provided by quantum mechanical Floquet 
theory for periodically forced systems, which has repeatedly been found
useful in the analysis of solid-state devices driven by external 
forces~\cite{WagnerZwerger97,GrifoniHaenggi98,LiReichl00,MartinezEtAl02}.
The material is organized as follows: Secs.~\ref{sec:single} and
\ref{sec:force} briefly provide the required background on localization 
at isolated defects, and on the influence of a periodic external force 
on a defect state. In Sec.~\ref{sec:split} an analytical expression for
the energy splitting between two defect states is derived, and generalized 
to describe the relevant quasienergy splitting when the periodic force
is turned on. Sec.~\ref{sec:twodefs} explains in detail the strategy for
achieving controlled population transfer between the two defects. 
Conclusions are drawn in the final Sec.~\ref{sec:conclusion}.

\section{Localization at a single defect}
\label{sec:single}

A single particle on a one-dimensional, infinite, regular lattice with matrix 
elements connecting neighboring sites only, as is appropriate in the 
tight-binding limit, is described by the Hamiltonian
\begin{equation}
\label{eq:h0}
   \hat{H}_0
   =
   -\frac{W}4 \sum_{\ell = -\infty}^{\infty}
   \left\{\ket{\ell}\bra{\ell+1} +  \ket{\ell+1}\bra{\ell}\right\}\,,
\end{equation}   
where $\ket{\ell}$ is the Wannier state localized at the $\ell$-th site, 
adopting the normalization $\bra{\ell'}\ell\rangle=\delta_{\ell',\ell}$.
Denoting the lattice constant by~$d$, its eigenstates
\begin{equation}
\label{eq:chi}
   \left|\chi_k\right>
   =
   \sum_{\ell = -\infty}^{\infty} \re^{\ri k\ell d}\ket{\ell}
\end{equation}
are extended Bloch waves, labeled by the wave number~$k$. The hopping 
matrix elements $-W/4$ in Eq.~(\ref{eq:h0}) have been chosen such that
the energy dispersion reads 
\begin{equation}
\label{eq:band}
   E(k) = -\frac{W}2\cos(kd)\,,
\end{equation}
corresponding to a band of width $|W|$. For positive $W$, its minimum lies at 
$k = 0$. In order to introduce a defect into the ideal system~(\ref{eq:h0}),
the on-site energy at the site~$\gamma$ is now altered by an amount~$\nu$, 
giving rise to the perturbation   
\begin{equation}
\label{eq:def}
   \hat{V}_{\rm r} = \ket{\gamma}\nu\bra{\gamma}\;,\quad \gamma>0 \;.
\end{equation}
The single-defect Hamiltonian
\begin{equation}
\label{eq:h1}
   \hat{H} = \hat{H}_0 + \hat{V}_{\rm r}
\end{equation}
then admits a localized state with energy
\begin{equation}
\label{eq:E_0}
   E_0=
   p\,\frac W2 \sqrt{\frac{4\nu^2}{W^2}+1}\;,
\end{equation}
where
\begin{equation}
\label{eq:defp}
   p=\left\{\begin{array}{r@{\quad:\quad}l}
     1&\frac{\nu}{W}>0\\
    -1&\frac{\nu}{W}<0\;,
\end{array}\right.
\end{equation}
so that $E_0$ falls either above (for $\nu/W > 0$) or below (for $\nu/W < 0$) 
the energy band~(\ref{eq:band}). The probability to find the particle at the 
$\ell$-th site, when it is bound by the defect, is given by 
\begin{equation}
\label{eq:loc}
{p}_{\ell}
= \frac{2\left|\nu\right|}{\sqrt{4\nu^2+W^2}}
   \left(\sqrt{\frac{4\nu^2}{W^2}+1}
   -2\left|\frac{\nu}W\right|\right)^{2\left|\ell-\gamma\right|}\;,
\end{equation}
as can be derived with the help of  resolvent operator techniques (see, 
{\em e.g.\/}, Ref.~\onlinecite{HoneHolthaus93}). In the following, knowledge of 
the amplitudes $a_\ell$ themselves will be required: As shown in the Appendix~\ref{sec:eigen}, the eigenstate
corresponding to the defect energy~(\ref{eq:E_0}) reads
\begin{equation}
   \ket{\psi_0}=\sum_{\ell=-\infty}^{\infty}a_{\ell}\ket{\ell}
\label{eq:psi0}
\end{equation}   
with
\begin{equation}   
   a_{\ell}={\cal N}(-p)^{\ell-\gamma}\left(x_-\right)^{\left|\ell-\gamma\right|}\;,
\label{eq:al}
\end{equation}
where $\cal N$ is the normalization constant, and the auxiliary quantities
\begin{equation}
\label{eq:x_pm}
x_{\pm} = \sqrt{\frac{4\nu^2}{W^2}+1}\pm 2\left|\frac{\nu}W\right|
\end{equation} 
have been introduced. They obey $x_-x_+=1$, and $x_-<x_+$. The fact that
the wave function~(\ref{eq:psi0}) with the amplitudes~(\ref{eq:al}) indeed
is normalizable for~$\frac{\nu}W\ne0$, and hence describes a localized state, 
follows immediately from $0<\left|x_-\right|<1$.

\section{Localization control by forcing}
\label{sec:force}

If an oscillating electric field is applied to the system, linearly polarized
along the direction of the lattice and with amplitude~$F$, the interaction
is modeled by 
\begin{equation}
\label{eq:force}
   \hat{H}_{\rm int}(t) 
   = 
   e F d \cos(\omega t)
   \sum_{\ell}\left|\ell\right>\ell\left<\ell\right|\;, 
\end{equation}
where $e$ is the particle's charge. Since the total Hamiltonian
$\hat{H}_0 + \hat{V}_{\rm r} + \hat{H}_{\rm int}(t)$ then is periodic 
in time, with period $T = 2\pi/\omega$, the Floquet 
theorem~\cite{Shirley65,Zeldovich67,Ritus67} asserts that there is a 
complete system of wave functions of the form
\begin{equation}
   \left|\psi(t)\right> = \re^{-\ri\varepsilon t/\hbar}\left|u(t)\right>\;,
\end{equation}
where $\left|u(t)\right>$ is a $T$-periodic function. The quantity 
$\varepsilon$ is  called ``quasienergy'', in analogy to the quasimomentum 
in solid state physics. 

If there is no defect, the quasienergies for a periodically driven particle 
in a tight-binding lattice with nearest-neighbor coupling 
read~\cite{Holthaus92,HolthausHone93} 
\begin{equation}
   \varepsilon(k) = -\frac{W}{2} 
   {\rm J}_0\left(\frac{eFd}{\hbar\omega}\right)\cos(kd) 
   \quad \bmod \hbar\omega \; ,
\end{equation}
so that the bandwidth is effectively quenched according to the zero-order
Bessel function ${\rm J}_0$: 
\begin{equation}
\label{eq:qeband}
   W_{\rm eff} = W{\rm J}_0\left(\frac{eFd}{\hbar\omega}\right)\;.
\end{equation}
At the zeros of ${\rm J}_0$, the band ``collapses''. This band collapse 
manifests itself as dynamical localization of the driven 
particle~\cite{DunlapKenkre86}, an effect which should be observable in 
far-infrared driven semiconductor superlattices~\cite{MeierEtAl95}. 

Without periodic forcing, the strength of a defect of the form~(\ref{eq:def}) 
is determined not by the on-site energy~$\nu$ alone, but rather by the ratio 
$\nu/W$: The larger $|\nu/W|$, the shorter is the localization length of the 
defect state, as witnessed by Eq.~(\ref{eq:loc}). This leads to the conjecture
that in the presence of periodic forcing the defect Floquet state again is
described by Eq.~(\ref{eq:loc}), but with $\nu/W$ replaced by 
$\nu/W_{\rm eff}$, so that the localization length of the defect state should 
become strongly dependent on the amplitude of the forcing. In particular,
when $eFd/(\hbar\omega)$ approaches a zero of ${\rm J}_0$, the state should 
be confined entirely to the defect site, if only $\nu \neq 0$.

\begin{figure}
\centerline{\includegraphics[width = 0.7\linewidth, angle=0]{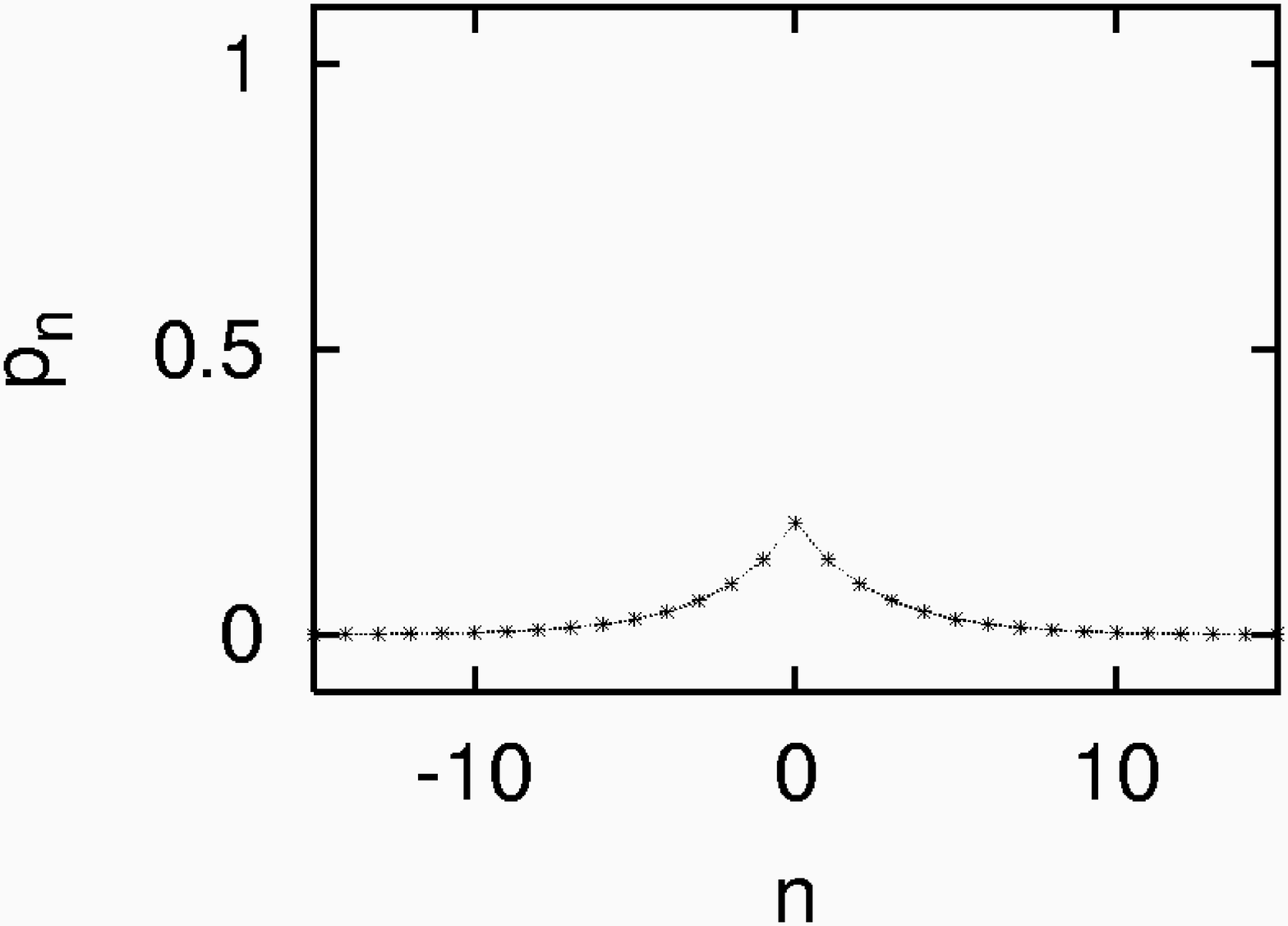}}
\centerline{\includegraphics[width = 0.7\linewidth, angle=0]{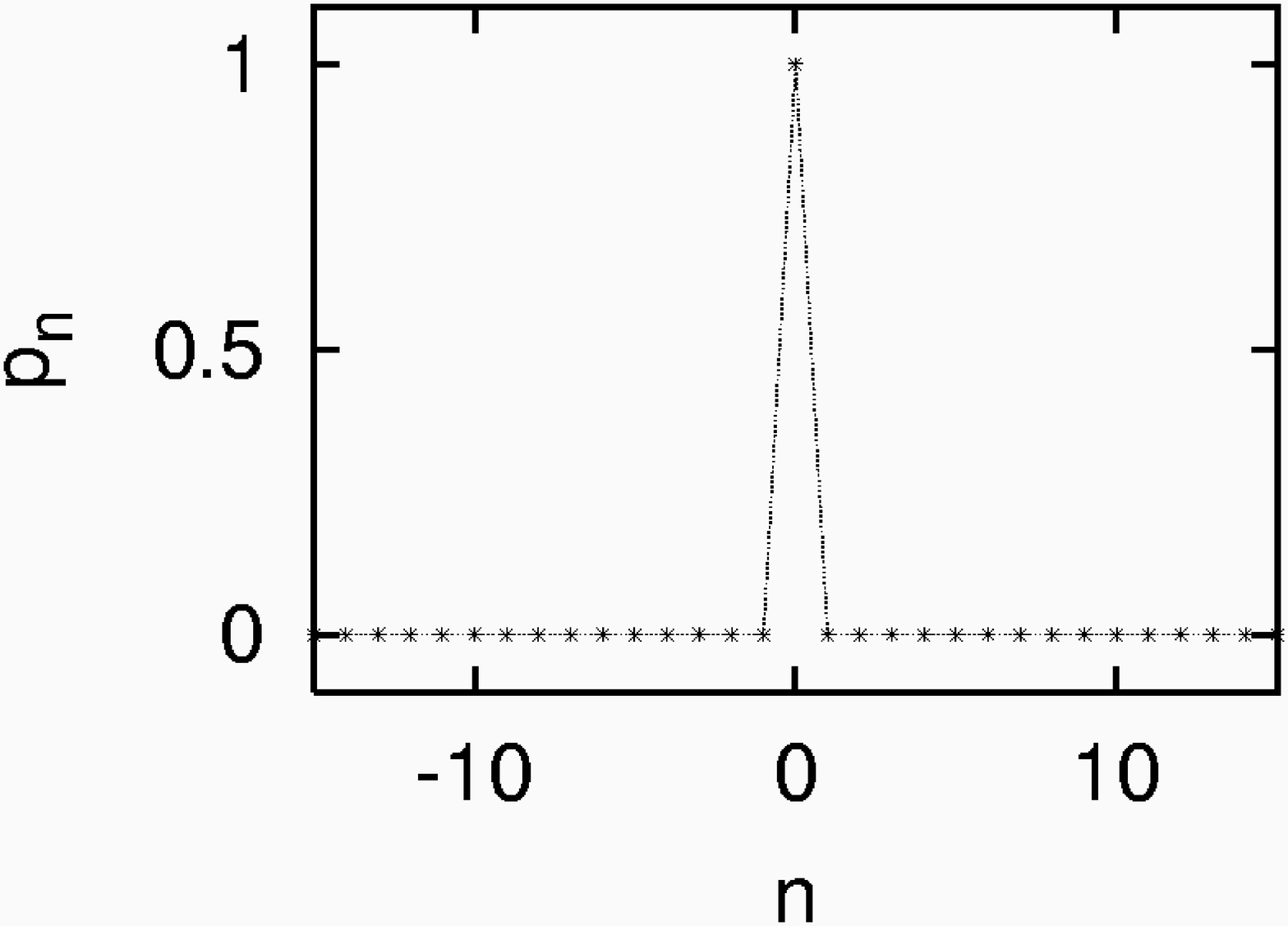}}
\caption[d]{Occupation probabilities $p_n$ for a defect state in a 
        one-dimensional lattice~(\ref{eq:h0}) with a defect of the 
        form~(\ref{eq:def}) at the site $n = 0$; the defect strength is 
        $\nu/W = 0.1$. The upper figure shows the state in the absence of 
        periodic forcing; it is exponentially localized according to 
        Eq.~(\ref{eq:loc}). The lower figure shows the defect Floquet
        state when the system is driven with frequency $\hbar\omega/W = 7.5$ 
        and amplitude $eFd/(\hbar\omega) = 2.4048$, corresponding to the 
        first zero of ${\rm J}_0$. Since the effective band 
        width~(\ref{eq:qeband}) is zero here, the defect state is localized 
        almost entirely at the defect site.}   
\label{fig:defect}
\end{figure}

As shown in Ref.~\onlinecite{HoneHolthaus93}, this conjecture indeed is correct in 
the high-frequency regime, where $\hbar\omega$ is significantly larger than
the bare band width $|W|$. This is illustrated in Fig.~\ref{fig:defect}
for a defect with strength $\nu/W = 0.1$: The upper panel shows the
occupation probabilities $p_n = |a_n|^2$ for a defect located at the
site $n = 0$ in the absence of the periodic force; the discrete values have 
been connected by lines to guide the eye. The lower panel shows the state
under the influence of a periodic force with high frequency 
$\hbar\omega/W = 7.5$ and scaled amplitude 
$eFd/(\hbar\omega) = j_{0,1} \simeq 2.4048$,
equal to the first zero of ${\rm J}_0$: As expected, the state now is
confined almost entirely to the defect site. Hence, the extension of the 
defect state is governed by the driving force's amplitude. This effect will 
be exploited in Sec.~\ref{sec:twodefs} to control the population transfer 
between two communicating defects.

\section{Energy splitting and quasienergy splitting for communicating defects}
\label{sec:split}

In the following, the real amplitudes $a_\ell$ for the state bound by
a single defect placed at a site $\gamma > 0$ are normalized such that
\begin{equation}
\label{eq:normdef}
   \sum_{\ell=1}^{\infty} \left|a_{\ell}\right|^2 = 1 \; ,
\end{equation}
which implies that the normalization constant ${\cal N}$ in 
Eq.~(\ref{eq:al}) is now given by
\begin{eqnarray}
   {\cal N} &=& \left({{\sum_{i=1}^{\gamma}x_-^{2(\gamma-i)}+
   \sum_{i=\gamma+1}^{\infty}x_-^{2(i-\gamma)}}}\right)^{-1/2}
\nonumber \\
   &=& \sqrt{\frac{1-x_-^2}{1+x_-^2-x_-^{2\gamma}}} \; .
   \label{eq:normerg} 
\end{eqnarray}
This defect state obeys the Schr\"odinger equation 
\begin{equation}
\label{eq:E0}
   \left( \hat{H}_0 + \hat{V}_{\rm r} \right)
   \left|\psi_0\right> = E_0\left|\psi_0\right> \; .
\end{equation}
Next, a second, identical defect is introduced into the left half of the 
lattice at the site labeled $-\gamma$, as described by
\begin{equation}
\label{eq:V_l}
   \hat{V}_{\rm l} = \ket{-\gamma}\nu\bra{-\gamma} \;,\quad \gamma>0\; .
\end{equation}
The two defects then carry two localized states $\left|\psi_{1}\right>$ 
and $\left|\psi_{2}\right>$, obeying the eigenvalue equations   
\begin{eqnarray}
\label{eq:E1}
   \left(\hat{H}_0 + \hat{V}_{\rm r} + \hat{V}_{\rm l} \right)
   \left|\psi_{1}\right> 
   &=& E_{1}\left|\psi_{1}\right> \; , \\
  \left(\hat{H}_0 + \hat{V}_{\rm r} + \hat{V}_{\rm l} \right)
   \left|\psi_{2}\right> 
   &=& E_{2}\left|\psi_{2}\right> \; .  
\end{eqnarray}
By symmetry, good approximations to these two defect states are given by
the even and odd linear combinations
\begin{equation}
   \left|\psi_{1,2}\right> 
   = 
   \sum_{\ell=-\infty}^{\infty}
   \frac 1{\sqrt{2}}\left(a_{\ell}\pm a_{-\ell}\right) 
   \left|\ell\right> \; .
\end{equation}
From these, auxiliary wave functions
\begin{eqnarray}
\label{eq:tilde}
   \widetilde{\left|{\psi}_0\right>} 
   &\equiv&
   \sum_{\ell=1}^{\infty}a_{\ell}\left|\ell\right>
\\ \mbox{and} \quad  
   \widetilde{\left|{\psi}_1\right>} 
   &\equiv&
   \sum_{\ell=1}^{\infty}
   \frac 1{\sqrt{2}}\left(a_{\ell}+ a_{-\ell}\right) 
   \left|\ell\right> 
\end{eqnarray}
are defined, which have nonvanishing amplitudes in the right half of the 
lattice only. Forming the scalar product of Eq.~(\ref{eq:E0}) with 
$\widetilde{\left<\psi_1\right|}$ then gives
\begin{equation}
   \widetilde{\left<\psi_1\right|} 
   \hat{H}_0 + \hat{V}_{\rm r}
   \left|\psi_0\right> 
   = E_0\widetilde{\left<\psi_1\right|} \left.\! \psi_0\right> \; ; 
\end{equation}
forming that of Eq.~(\ref{eq:E1}) with $\widetilde{\left<\psi_0\right|}$ 
leads to 
\begin{equation}
   \widetilde{\left<\psi_0\right|} 
   \hat{H}_0 + \hat{V}_{\rm r} 
   \left|\psi_1\right> 
   = E_1\widetilde{\left<\psi_0\right|} \left.\! \psi_1\right> \; ,
\end{equation}
since 
$\widetilde{\left<\psi_0\right|} \hat{V}_{\rm l}
= 0$ (see Eqs.~(\ref{eq:V_l}) and (\ref{eq:tilde})).
By definition,
$\widetilde{\left<\psi_1\right|}\left.{\psi}_0\right> =
\widetilde{\left<\psi_0\right|}\left.{\psi}_1\right>$. Hence, subtracting 
the above two equations yields 
\begin{equation}
   \left( E_0 - E_1 \right) 
   \widetilde{\left<\psi_1\right|} \left.\! \psi_0\right> = 
   \widetilde{\left<\psi_1\right|} \hat{H}_0 \left|\psi_0\right> - 
   \widetilde{\left<\psi_0\right|} \hat{H}_0 \left|\psi_1\right> \; .
\end{equation}
It is now stipulated that the localization of the state $\left|{\psi}_0\right>$
around the site $\gamma > 0$ be sufficiently strong that its amplitudes
in the left half of the lattice are negligible. One then has
\begin{equation}
   \widetilde{\left<\psi_1\right|}\left.{\psi}_0\right>\simeq
   \frac1{\sqrt{2}}\widetilde{\left<\psi_0\right|}\left.{\psi}_0\right> 
   = \frac1{\sqrt{2}} \; , 
\end{equation}
leading to
\begin{eqnarray}
   E_0-E_1&\simeq& 
   -\frac W{4}\sum_{\ell=1}^{\infty} \Big\{
   \left(a_{\ell}\!+\!a_{-\ell}\right)a_{\ell+1}
   +\left(a_{\ell}\!+\!a_{-\ell}\right)a_{\ell-1}\Big\} 
\nonumber \\
    & &+\frac W{4}\sum_{\ell=1}^{\infty} \Big\{
    a_{\ell}\left(a_{\ell+1}\!+\!a_{-\ell-1}\right)  
\nonumber \\ & & \qquad \qquad \qquad
    +a_{\ell}\left(a_{\ell-1}\!+\!a_{-\ell+1}\right) \Big\}
\nonumber \\
    &= &-\frac W{4}\sum_{\ell=1}^{\infty} 
    \Big\{ \big(a_{-\ell}a_{\ell+1}-a_{-\ell+1}a_{\ell} \big) 
\nonumber \\ & & \qquad \qquad 
    + \big(a_{-\ell}a_{\ell-1}-a_{-\ell-1}a_{\ell}\big) \Big\} \; .
\end{eqnarray}
These telescope series can be summed immediately ($\sum_{n=1}^N\left(b_n-b_{n-1}\right)=b_N-b_0$ and
$\lim_{\ell\rightarrow\infty}a_{\ell}=0$), resulting in
\begin{equation}
   E_0-E_1=\frac W{4} \left(a_1-a_{-1}\right)a_0 \; .
\end{equation}
In the same manner, one also derives
\begin{equation}
    E_2-E_0=\frac W{4} \left(a_1-a_{-1}\right)a_0 \; .
\end{equation}
Summing these two equations finally leads to a surprisingly simple
expression for the splitting of the energies associated with the
two defects: 
\begin{equation}
\label{eq:fin}
   E_2-E_1=\frac W{2} \left(a_1-a_{-1}\right)a_0 \; .
\end{equation}
Since $\left(a_1-a_{-1}\right)/2$ can be taken as the discrete derivative
of the wave function at the site $\ell = 0$, the energy splitting is
determined by the product of the wave function itself and its derivative
halfway between the two defects, {\em i.e.\/}, by the current prevailing  
there. Thus, this Eq.~(\ref{eq:fin}) constitutes a discrete analog of
Herring's formula, which describes the tunneling splitting for wave 
functions in double well potentials~\cite{LL,Herring62,Gutzwiller90}. 
In view of Eqs.~(\ref{eq:al}) and (\ref{eq:normerg}), one then has 
\begin{eqnarray}
   \Delta E &\equiv& E_2-E_1 
\nonumber\\
   &=&p\frac W2 \frac{1-x_-^2}{1+x_-^2-x_-^{2\gamma}}x_-^{2\gamma}
   \left(x_--x_-^{-1}\right) \; ,
\end{eqnarray}
or
\begin{equation}
\label{eq:relative}
   \frac{\Delta E}{E_0} = -\frac{\left(1-x_-^2\right)^2x_-^{2\gamma-1}}
   {\left(1+x_-^2-x_-^{2\gamma}\right) \sqrt{\frac{4\nu^2}{W^2}+1}}\;. 
\end{equation}

As with a single defect, the energy eigenvalues are replaced by the
corresponding quasienergies in the presence of time-periodic 
forcing, and the energy splitting turns into a quasienergy splitting.
The above result~(\ref{eq:relative}), with $W$ replaced by $W_{\rm eff}$
according to Eq.~(\ref{eq:qeband}), should also be a good approximation
to the quasienergy splitting if the driving frequency is sufficiently high.

\begin{figure}
\centerline{\includegraphics[width = 0.7\linewidth, angle=0]{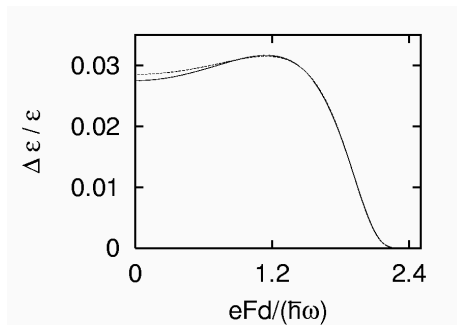}}
\caption[d]{Relative splitting of the quasienergies in a periodically driven 
        lattice with two identical defects of strength $\nu/W = 0.1$ situated 
        at $\gamma = \pm 3$, under the influence of forcing with frequency 
        $\hbar\omega/W = 7.5$. The full line shows the exact, numerically 
        obtained data. The dashed line results from the Herring-type 
        approximation~(\ref{eq:relative}), with the bandwidth~$W$ replaced 
        by the effective band width~(\ref{eq:qeband}).} 
\label{fig:delta}
\end{figure}

In order to check this hypothesis, the time-dependent Schr\"odinger equation
for the periodically forced two-defect system has been solved numerically, 
and the quasienergies for the localized states have been obtained. The 
results for two defects at $\gamma = \pm 3$, with $\nu/W = 0.1$ and 
$\hbar\omega/W = 7.5$, are plotted in Fig.~\ref{fig:delta} as functions of 
the scaled amplitude $eFd/(\hbar\omega)$. As can be seen, the agreement 
between the analytical approximation and the exact numerical data becomes 
excellent when $eFd/(\hbar\omega) > 1$.

\section{Controlled population transfer}
\label{sec:twodefs}

It is assumed now that initially, at time $t = 0$, the particle is 
localized at one of the two defects, and the periodic force is present. 
Denoting the two Floquet functions associated with the defects at the 
given driving amplitude~$F$ by $\left|u_{1,2}^{F}(t)\right>$, and their
quasienergies by $\varepsilon^{F}_{1,2}$, the initial state is given by a
superposition      
\begin{equation}
\label{eq:ini}
   \left|\psi(0)\right> =  \frac1{\sqrt{2}}\big( \left|u_1^{F}(0)\right> 
   \pm \left|u_2^{F}(0)\right> \big) \; .
\end{equation} 
Under the influence of forcing with constant amplitude, this state
evolves in time according to   
\begin{eqnarray}
   \left|\psi(t)\right> & = & \frac1{\sqrt{2}}\Big( \left|u_1^{F}(t)\right> 
   \pm \left|u_2^{F}(t)\right> \exp(-\ri \Delta\varepsilon^{F}t/\hbar) \Big)
\nonumber \\ & & \times
   \exp(-\ri \varepsilon_1^Ft/\hbar) \; ,
\end{eqnarray} 
where
\begin{equation}
   \Delta \varepsilon^F = \varepsilon_2^F - \varepsilon_1^F
\end{equation}
denotes the quasienergy splitting. Hence, the particle is coherently
oscillating between the two defects; the transfer time $T_{\rm trans}$,
after which the particle will be found at the other defect, is given by
\begin{equation}
   T_{\rm trans} = \frac{\pi \hbar}{\Delta \varepsilon^F}
\end{equation}
and thus depends on the driving amplitude~$F$.

When the amplitude changes sufficiently slowly in time, the system responds 
in an adiabatic manner~\cite{BreuerHolthaus89}. Hence, under the influence 
of a slowly varying amplitude~$F(t)$, the initial state~(\ref{eq:ini}) follows 
the instantaneous Floquet states and evolves into
\begin{eqnarray}
   \left|\psi(t)\right> & = & \frac1{\sqrt{2}}\left[ \left|u_1^{F(t)}(t)\right>\right.
\nonumber \\ & &  \left.
   \pm \left|u_2^{F(t)}(t)\right> 
   \exp\left(-\frac{\ri}{\hbar} \int_0^t \! \rd \tau \,  
   \Delta\varepsilon^{F(\tau)}\right) \right]
\nonumber \\ & & \times
   \exp\left(-\frac{\ri}{\hbar} \int_0^t \! \rd \tau \,
   \varepsilon_1^{F(\tau)}\right) \; .  
\end{eqnarray} 
This implies that the transfer time from one defect to the other now
is given by the relation
\begin{equation}
\label{eq:int}
   \frac{1}{\hbar} \int_0^{T_{\rm trans}} \! \rd \tau \, 
   \Delta\varepsilon^{F(\tau)} = \pi \; , 
\end{equation}
which constitutes an immediate analog of the $\pi$-pulse-condition known
from two-level systems~\cite{AllenEberly87}. This is the physics which will 
now be exploited for coherent control of population transfer between two 
communicating defects. 

To this end, the driving amplitude $F(t)$ is shaped such that one has
$eFd/(\hbar\omega) = j_{0,1}$ for $t\le 0$, so that the defect Floquet
states are confined to their respective sites, and the communication 
between the two defects is effectively disrupted. Then the amplitude is
adiabatically lowered such that the defect states start to overlap
significantly, and the particle initially tied to one defect oscillates 
to the other. If the amplitude then rises again and reaches the ``collapse'' 
value
\begin{equation}
   F_{\rm collapse} = j_{0,1} \frac{\hbar\omega}{ed} 
\end{equation}
at $t = T_{\rm pulse}$, and is kept constant thereafter, the particle
has been transferred to the final state, and will stay there.
The time~$T_{\rm pulse}$ is chosen such that
\begin{equation}
\label{eq:intphi}
   \frac{1}{\hbar} \int_0^{T_{\rm pulse}} \! \rd \tau \, 
   \Delta\varepsilon^{F(\tau)} = \varphi\; ; 
\end{equation}
to obtain a transfer from one defect to the other, again~$\varphi=\pi$ has to be chosen.

For a matter-of-principle demonstration of this scenario, one may
employ the envelope function  
\begin{eqnarray}
\label{eq:env}
   F(t) & = & F_{\rm collapse}
\\ &\times&
   \frac{\exp\left(-\frac {t^2}{2T_{\rm ramp}^2}\right) + a +
   \exp\left(-\frac{\left(t-T_{\rm pulse}\right)^2}{2T_{\rm ramp}^2}\right)}  
   {a+1 + \exp\left(-{\frac {T_{\rm pulse}^{2}}{2T_{\rm ramp}^2}}\right)} 
\nonumber
\end{eqnarray}
for $0 \le t \le T_{\rm pulse}$, where $T_{\rm ramp}$ quantifies the 
characteristic time interval during which the amplitude is ramped down 
and up again, with $T \ll T_{\rm ramp} \ll T_{\rm pulse}$ being understood.
The parameter $a$ has been introduced in order to allow for a nonvanishing
amplitude at intermediate times, at $t \approx T_{\rm pulse}/2$.

\begin{figure}[h]
\centerline{\includegraphics[width = 0.7\linewidth, angle=0]{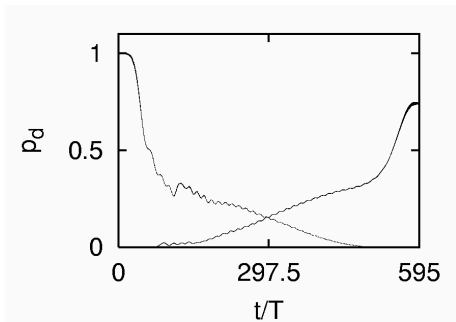}}
\caption[d]{Incomplete population transfer resulting from fast, nonadiabatic
        changes of the amplitude function. There are two defects at
        $\gamma = \pm 3$ with strength $\nu/W = 0.1$; the forcing frequency
        is $\hbar\omega/W = 7.5$. The envelope function is  given by
        Eq.~(\ref{eq:env}), with $a = 1.5$, $T_{\rm ramp} = 50\,T$, and
        $T_{\rm pulse} = 595\,T$. Plotted are the probabilities to find
        the particle at the two defect sites.}
\label{fig:versuch}
\end{figure}

\begin{figure}[h]
\centerline{\includegraphics[width = 0.7\linewidth, angle=0]{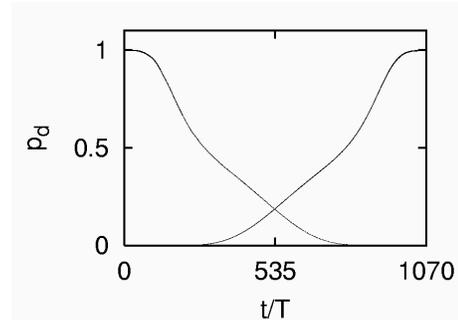}}
\caption[d]{Adiabatic population transfer at communicating defects:
        Now $T_{\rm ramp} = 200\,T$, and $T_{\rm pulse} = 1070\,T$; the other
        parameters are as in Fig.~\ref{fig:versuch}.}   
\label{fig:trans}
\end{figure}

The above considerations rely heavily on the adiabatic principle; if the
parameter variation proceeds too fast, complete population transfer is
not achieved. This is illustrated in Fig.~\ref{fig:versuch} for a system 
with defect parameters $\gamma = \pm3$ and $\nu/W = 0.1$, subjected to
forcing with frequency $\hbar\omega/W = 7.5$ and the envelope 
function~(\ref{eq:env}), setting $a = 1.5$, $T_{\rm ramp} = 50\,T$, and
$T_{\rm pulse} = 595\,T$. Under these conditions, the transfer remains
incomplete; slight oscillations visible in the occupation probabilities
of the defect sites indicate non-adiabatic dynamics. However, 
Fig.~\ref{fig:trans} demonstrates that the desired result is obtained
when the time scales are prolonged: With $T_{\rm ramp} = 200\,T$ and 
$T_{\rm pulse} = 1070\,T$, one has practically complete transfer.

\begin{figure}
\centerline{\includegraphics[width = 0.7\linewidth, angle=0]{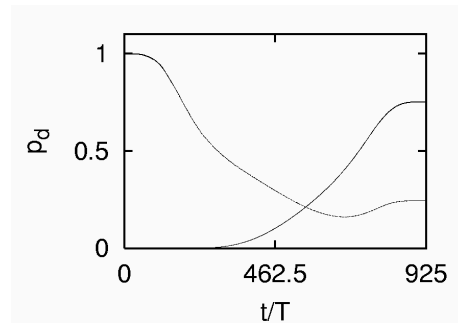}}
\caption[d]{Probabilities to find the particle at the two defects for 
        $T_{\rm ramp} = 200\,T$ and $T_{\rm pulse} = 925\,T$; the other
        parameters are as in Fig.~\ref{fig:versuch}. This choice of
        parameters aims at producing a coherent, Schr\"odinger cat-like
        state, with probabilities of $25\%$ and $75\%$ for finding the
        particle at the defects. Arbitrary other final probability ratios
        can be achieved as well.}
\label{fig:super}
\end{figure}

It is, of course, also possible to employ the method outlined here to 
prepare the particle in superpositions of defect states. For instance, if 
one chooses $T_{\rm pulse}$ such that the phase integral in Eq.~(\ref{eq:int})
yields $\pi/2$, rather than $\pi$, the resulting state describes a particle 
which, after initially being localized at a single defect, is eventually 
found with equal probability on either one. In the same manner, any desired 
probability ratio can be obtained; Fig.~\ref{fig:super} shows an example where 
the particle remains with a probability of $1/4$ at the initial defect, and is 
found with a probability of $3/4$ at the other one, after forcing according 
to Eq.~(\ref{eq:env}).

\section{Conclusion}
\label{sec:conclusion}

It has been shown in this paper that a periodic force can drastically
alter the energy splitting associated with the two states bound by 
two identical defects in a one-dimensional tight-binding lattice. In
the presence of the force, the band width~$W$ entering the energy 
splitting~(\ref{eq:relative}) has to be replaced by the effective
width~(\ref{eq:qeband}), so that the splitting can be monitored within 
wide ranges, and even be completely suppressed. Thus, the times required
for coherent population exchange between the defects can be varied over
several orders of magnitude by adjusting the amplitude of the force.

The strategy employed here to give a matter-of-principle illustration 
of coherent control of population transfer relies on the adiabatic 
principle, and thus is restricted to forces with slowly varying amplitudes. 
It appears possible to overcome this restriction: Utilizing techniques 
developed for the coherent control of 
molecules~\cite{JudsonRabitz92,AssionEtAl98}, it might be possible 
to design even rapidly changing envelopes which effectuate a guided 
transport of a particle from one defect to the other, or to create 
superposition states with well-defined weights. This might open up 
new perspectives for the design of quantum logical devices.     

\acknowledgments
I would like to thank M.~Holthaus for his continuous support and insightful discussions.

\appendix

\section{\label{sec:eigen}Eigenfunction for a single defect}

In order to demonstrate that $\ket{\psi_0}$ as defined by Eqs.~(\ref{eq:psi0})
and~(\ref{eq:al}) indeed is an eigenfunction of the single-defect 
Hamiltonian~(\ref{eq:h1}) with the energy eigenvalue~(\ref{eq:E_0}), it is 
helpful to invoke the definition~(\ref{eq:defp}), and to write
$\nu = W\frac{\nu}W = W p\left|\frac{\nu}W\right|$.
One then obtains
\begin{eqnarray}
   \hat{H}\ket{\psi_0}=&& 
   \ket{\gamma}pW\left|\frac{\nu}{W}\right|\bra{\gamma}{\psi_0}\rangle
\\
   &-&\frac{W}4 \sum_{\ell = -\infty}^{\infty}
   \left\{\ket{\ell}\bra{\ell+1} +  \ket{\ell}\bra{\ell-1}\right\}\ket{\psi_0}
\nonumber\\
   =&& pW\left|\frac{\nu}{W}\right|a_{\gamma}\ket{\gamma}
   -\frac{W}4 \sum_{\ell = -\infty}^{\infty}
   \left(a_{\ell-1}+a_{\ell+1}\right)\ket{\ell}
\nonumber\;.\nonumber
\end{eqnarray}
Singling out the defect site, one has
\begin{eqnarray}
   \hat{H}\ket{\psi_0}=&&
   \left(pW\left|\frac{\nu}{W}\right|a_{\gamma}
   -\frac{W}4\left(a_{\gamma-1}+a_{\gamma+1}\right)\right)\ket{\gamma} 
\nonumber\\&   
   -&\frac{W}4 \sum_{\genfrac{}{}{0pt}{1}{\ell = 
   -\infty}{\ell\ne\gamma}}^{\infty}\left(a_{\ell-1}
   +a_{\ell+1}\right)\ket{\ell}\;.
\end{eqnarray}
Utilizing the explicit expression~(\ref{eq:al}) for the 
amplitudes~$a_{\ell}$, this leads to 
\begin{eqnarray}
   \hat{H}\ket{\psi_0}=&& 
   \left(pW\left|\frac{\nu}{W}\right|+p\frac{W}2x_-\right)a_{\gamma}\ket{\gamma}
\nonumber\\
&-&\frac{W}4\sum_{\genfrac{}{}{0pt}{1}{\ell = -\infty}{\ell\ne\gamma}}^{\infty}
   (-p)(x_-^{-1}+x_-)a_{\ell}\ket{\ell}
\nonumber\\
   =&&p\frac{W}2\sqrt{\frac{4\nu^2}{W^2}+1}\,a_{\gamma}\ket{\gamma} 
\nonumber\\
   &+&p\frac{W}4 \sum_{\genfrac{}{}{0pt}{1}{\ell = 
   -\infty}{\ell\ne\gamma}}^{\infty}(x_-^{-1}+x_-)a_{\ell}\ket{\ell}\;.
\end{eqnarray}
According to the definition~(\ref{eq:x_pm}), one has $x_-^{-1} = x_+$.
Hence, $x_-^{-1}+x_- = x_+ + x_- =2\sqrt{\frac{4\nu^2}{W^2}+1}$, which implies
\begin{equation}
   \hat{H}\ket{\psi_0} =
   p\frac{W}2\sqrt{\frac{4\nu^2}{W^2}+1}
   \sum_{\ell = -\infty}^{\infty}a_{\ell}\ket{\ell}\;.
\end{equation}
This proves the assertion.

\end{document}